\documentstyle[amsfonts,12pt]{article}
\RequirePackage{graphicx}
\RequirePackage{times}
\RequirePackage{courier}
\RequirePackage{mathptm}

\begin{document}


\title{Conventions in relativity theory and quantum mechanics}
\author{Karl Svozil\\
 {\small Institut f\"ur Theoretische Physik, University of Technology Vienna }     \\
  {\small Wiedner Hauptstra\ss e 8-10/136,}
  {\small A-1040 Vienna, Austria   }            \\
  {\small e-mail: svozil@tuwien.ac.at}}
\date{ }
\maketitle


\begin{abstract}
The conventionalistic aspects of physical world perception are reviewed
with an emphasis on the constancy of the speed of light in relativity theory
and the irreversibility of measurements in quantum mechanics.
An appendix contains a complete proof of Alexandrov's theorem
using mainly methods of affine geometry.
\end{abstract}

\section{Know thyself}
This inscription on the Oracle of Apollo at Delphi, Greece,
dates from 6th century B.C., and it is still of tremendous importance
today.
For we do not and never will see the world ``as is,'' but rather as we
perceive it.
And how we perceive the world is mediated by our senses which serve
as interfaces to the world ``out there'' (if any); but
not to a small extend also by what we project onto it.
Conventions are projections which we have to adopt in order to be able
to cope with the phenomena ``streaming in'' from the senses.
Conventions are a necessary and indispensable part of
operationalizable\footnote{
In what follows we shall adopt Bridgman's concept of ``operational''
as one of quite simple-minded experimental testability,
even in view of its difficulties which this author approached later on
\cite{bridgman27,bridgman,bridgman36,bridgman50,bridgman52}.}
phenomenology and tool-building.
There is no perception and intervening without conventions.
They lie at
the very foundations of our world conceptions.
Conventions serve as a sort of ``scaffolding'' from which
we construct our scientific worldview.
Yet, they are so simple and almost self-evident that they are hardly
mentioned and go unreflected.

To the author, this unreflectedness and unawareness of conventionality
appears to be the biggest problem related to conventions,
especially if they are mistakenly considered as physical ``facts''
which are empirically testable.
This confusion between assumption and observational, operational fact
seems to be one of the biggest impediments for progressive research programs,
in particular if they suggest postulates which are based on conventions
different from the existing ones.

In what follows we shall mainly review and discuss conventions in the
two dominating theories of the 20th century: quantum mechanics and
relativity theory.

\section{Conventionality of the constancy of the characteristic speed}

Suppose two observers called Alice and Bob
measure space and time in two
coordinate frames.
Operationally their activities amount to the following.
They have constructed ``identical'' clocks and scales of ``equal''
length
which they have compared in the distant past; when Bob  lived
together with Alice.
Then they have separated.
Alice has decided to depart from Bob and, since then, is moving with
constant speed away from him.
How do Bob's and Alice's scales and clocks compare now?
Will they be identical, or will they dephase?

These are some of the questions which ``relativity'' theory deals with.
It derives its name from Poicar\'e's 1904 ``priciple of relativity''
stating that
\cite[p. 74]{miller-1998} {\em ``the laws of physical phenomena must be
the same
for a stationary observer as for an observer carried along in a uniform
translation; so that we have not and can not have any means of
discerning whether or not we are carried along in such a motion.''}
Formally, this amounts to the requirement of form invariance or
covariance of the physical equations of motion.

One of the seemingly mindboggling features of the theory of
relativity is the fact that simultaneity and even the time order of two
events needs no  longer be conserved.
It may indeed happen that Alice perceives the first event before the second,
while Bob perceives both events as happening at the same time; or even
the second event ahead of the first.
Simulaneity can only be defined ``relativ'' to a particular reference
frame. If true there, it is false in any different frame.

The first part of Einstein's seminal paper
\cite{ein-05} is dedicated to a detailed study of the
intrinsically operational procedures and methods which are necessary to
conceptualize the comparison of Alice's and Bob's reference frames.
This part  contains certain ``reasonable'' conventions
for defining clocks, scales,
velocities and in particular simultaneity,
without which no such comparison could ever be achieved.
These conventions appear to be rather evident and natural,
almost trivial, and yield a convenient
formalization of space-time transformations, but they are
nevertheless arbitrary.
The simultaneity issue has been much debated in the contemporary
discussions on conventionality  \cite{malamet,redhead-93,stanf-enc-syn}.

There is another element of conventionality present in relativity
theory which has gotten less attention \cite{svozil-relrel}.
It is the assumption of the constancy of the speed of light.
Indeed, the International System of units assumes light to be
constant.
It was
decided in 1983 by the General Conference on Weights and Measures that the
accepted value for the speed of light would be exactly 299,792,458
meters per
second. The meter is now thus  defined as the distance traveled by
light in a vacuum in 1/299,792,458 second, independent of the
inertial system.

Despite the obvious conventionality of the constancy of the speed of
light, many introductions to relativity theory present this proposition
not as a convention but rather as an important empirical finding.
Indeed, it is historically correct to claim that experiments like
the ones of Fizeau, Hoek and Michelson-Morley, which
produced a null result by attempting to measuring the earth's motion
against some kind of ``ether,'' preceded Einstein's special theory of
relativity.

But this may be misleading.
First of all, Einstein's major reason for introducing the Lorentz
transformation seems to have been the elimination of asymmetries
which appeared in the
electromagnetic formalism of the time but are not inherent in the
phenomena, thereby unifying electromagnetism.
Secondly, not too much consideration has been given to the possibility
that experiments like the one of Michelson and Morley may be
a kind of
``self-fulfilling prophesy,'' a circular, closed tautologic exercise.
If the very instruments
which should indicate a change in the velocity of light are themselves
dilated, then any dilation effect will be effectively nullified.
This possibility has already been imagined in the 18th century
by Boskovich \cite{bos} and was later put forward by
FitzGerald
\cite{FitzGerald2} (see also John Bell \cite{bell-sr1,bell-92}),
Lorentz, Poincar\'e  and others in the
context of the ether theory \cite{miller-1998}.\footnote{
In stressing the conventionality aspect of these effects,
the author would like to state that he does not want to promote
any ether-type theory, nor is he against any such attempts.}

But what is the point in arguing that the constancy of
the speed of light is a convention rather than an empirical finding?
Is this not a vain question; devoid of any operational testability?

The answer to this concern is twofold.
First, a misinterpretation might give rise to a doctrinaire and improper
preconception
of relativity theory by limiting the scope of its
applicability.
Indeed, as it turns out, for reasons mentioned below
\cite{svozil-relrel},
the special theory
of relativity is much more generally applicable as is nowadays appreciated.
It applies also to situations in which
the velocity of light needs not necessarily
be the highest possible limit velocity for signaling
and travel.
Secondly, it is not totally unreasonable to ask the following question.
What if one adopts a different convention by assuming a different
velocity than that of light to be the basis for frame generation?
Such a velocity may be anything, sub- but also superluminal. What will
be the changes to Alice's and Bob's frames, and how do these new
coordinates relate to the usual ``luminal'' frames?

These issues have been discussed by the author
\cite{svozil-relrel}  on the basis of a geometrical theorem by
Alexandrov \cite{alex1,alex2,alex3,alex-col} and Borchers and Hegerfeldt
\cite{borchers-heger} reviewed in \cite{benz,lester}
(see also previous articles \cite{svo5,svo-86}).
Alexandrov's theorem requires the convention that some speed  is
the same in Bob's and Alices's frames.
Furthermore, if two space-time points are different in Alice's frame,
then these points must also be mapped into different points in Bob's
frame and vice versa; i.e., the mapping must be one-to-one, a bijection.
It can be proven that under these conditions, the mapping relating
Alice's and Bob's
frames must be an affine Lorentz transformation, with some fundamental speed
playing
the role of light in the usual Lorentz transformations of relativity
theory.
The nontrivial geometric part of a proof uses the theorem of affine
geometry, which results in the linearity of the transformation
equations.
No Poicar\'e's 1904 ``priciple of relativity,'' no
relativistic form invariance or covariance is needed despite
the postulate or convention of equality of a single speed in all
reference frames.
The derivation uses geometry, not physics.
The Appendix contains a detailed derivation of Alexandrov's theorem which should be
conprehendible for a larger audience.

To repeat the gist: it is suggested that the signalling velocity
occuring in the Lorentz transformation is purely conventional.
This effectively turns the interpretation of relativity theory upside
down and splits it into two parts, one geometric and one physical, as
will be discussed next.

So where is the physics gone?
The claim of conventionality arouses suspicions.
The proper space-time transformations cannot be purely conventional or
even a matter of epistemology!
After all, the Michelson-Morley experiment and most of its various pre-
and successors actual yielded null results, which are valid physical
observations as can be.
The experimenters never explicitly acknowledged the conventionality of
the constancy of the speed of light and approved their instruments
according to these specifications.
Just on the contrary, they first assumed to measure the unequality
and anisotropy of the speed of light.
And what if Alice and Bob assume, say, the constancy of the speed of
sound instead of light?
Would the mere assumption change the reading of the instruments in a
Michelson-Morley experiment using sound instead of light?
This seems to be against all intuitions and interpretations and the
huge accumulated body of evidence.

The answer to these issues can be sketched as follows.
First of all, the physics is in the form invariance of the
electromagnetic equations under a particular type of Lorentz
transformations: those which contain the speed of electromagnetic
signals; i.e., light, as the invariant speed.
Thus, merely the convention of the constancy of the speed of light in
all reference frames yields the desirable relativistic covariance of the
theory of electromagnetism.
This is a preference which cannot be motivated by geometry or
epistemology; it is purely physical.

However, any such Lorentz transformation will result in a non-invariance
of the theory of sound or any other phenomena which are not directly
dominated by electromagnetism. There, an asymmetry will appear, singling
out a particular frame of reference from all the other ones.

Thus, one may speculate that the most efficient
``symmetric'' representation  of the physical laws is by
transformations which assume the convention  of the invariant signalling
velocity which directly reflects the phenomena involved.
For electromagnetic phenomena it is the speed of electromagnetic waves;
i.e., light.
For sound phenomena it is the speed of sound.
For gravity it is the speed of gravitational waves.
Thus the conventionality of the relativity theory not only relativizes
simultaneity
but must also reflect the particular class of phenomena; in particular
their transmission speed(s).
In that way, a general form invariance or covariance is achieved,
satisfying  Poicar\'e's 1904 ``priciple of relativity'' mentioned above,
which is not only limited to electromagnetism but is valid also for a
wider class of systems.

Secondly, it is not unreasonable to assume that in the particular
context of the Michelson-Morley and similar experiments,
all relevant physical system
parameters and instruments are governed by electromagnetic phenomena
and not by sound, gravity or something else.
Thus, although not explicitly intended,
the experiments are implicitly implementing the conventionality
of the constancy of light.
Of course, the experimenter could decide to
counteract the most natural way to gauge the instruments and measure
space and time differently than as suggested by the intruments.
For instance, one may adopt scales to measure space which are
anisotropic and velocity dependent.
But this would be a highly unreasonable, unconvenient and complicating
thing to do.

So, from a system theoretic standpoint,
the proper convention suggests itself
by the dominating type of interaction,
and only in this way corresponds
to a physical proposition.
The result is a generalized principle of relativity.

\section{Conventionality of quantum measurements}

In what follows the idea is put forward and reviewed that measurements
in  quantum mechanics are (at least in principle) purely conventional.
More precisely, it is purely conventional and subjective what exactly an
``observer'' calls ``measurement.''
There is no distinction between ``measurement'' and ordinary
(unitary) quantum evolution other that the particular interpretation
some (conscious?) agent ascribes to a particular process
\cite{svozil-2000interface}.
Indeed, the mere distinction between an ``observer'' and the ``object''
measured is purely conventional.
Stated pointedly. measurement is a subjective illusion.
We shall call this the ``no measurement'' interpretation of quantum
mechanics.

The idea that measurements, when compared to other processes (involving
entanglement), are nothing special, seems to be quite widespread among
the quantum physics community; but it is seldomly spelled out publicly
\cite{greenberger2,greenberger-talk-99}.
Indeed, the possibility to ``undo'' a quantum
measurement has been experimentally documented  \cite{hkwz},
while it is widely acknowledged that practical bounds to maintain
quantum coherence pose an effective upper limit on the possibility to
reconstruct a quantum state.
We shall not be concerned with this upper bounds, which does not seem to
reflect some deep physical truth but rather depends on
technology, financial commitments and cleverness on the experimenter's
side.

Rather, we shall discuss the differences between the two types of
time evolution which are usually postulated in quantum mechanics:
(i) unitary, one-to-one, isometric time
evolution inbetween
two measurements and (ii) many-to-one state reduction at the measurement.

Inbetween two measurements, the quantum state
undergoes a deterministic, unitary time evolution, which is reversible
and one-to-one. This amounts to arbitrary generalized
isometries---distance-preserving maps---in complex Hilbert space.
A discrete analogue of this situation is the permutation of
states.
An ``initial message'' is constantly being re-encoded.
As such an evolution is reversible, there is no principle reason why
any such evolution cannot be undone. (There may be insurmountable
practical obstacles, though.)

Any irreversible measurement is formally accompanied by a
state reduction or ``wave function collapse'' which is many-to-one.
Indeed, this many-to-oneness is the formal mathematical expression of irreversibility.

What is a measurement? Besides all else, it is associated with a
some sort of ``information''
transfer through a fictious boundary between some ``measurement
apparatus'' and the ``object.''
In the following we shall call this fictious boundary the ``interface.''
The interface has no absolute physical relevance but is purely
conventional.
It serves as a scaffolding to mediate and model the exchange.
In principle, it can be everywhere.
It is symmetric: the role of the
observer and the observed system is interchangeable and a
distinction is again purely conventional.

In more practical terms, it is mostly rather obvious what is the observer's
side. It is usually inhabited by a conscious experimenter and his
measurement device.
It should be also in most cases quite reasonable to define the
interface as the location where some agent serving as the experimenter
looses
control of one-to-onenness. This is the point where ``the quantum turns
classical.''
But from the previous discussion it should already be quite
clear that any irreversibility in no way reflects a general
physical principle but rather the experimenter's ability to reconstruct
previous states.
Another ``observer'' equipped with another technology (or just more
money) may draw very different interface lines.

Let me add one particular scenario for quantum information.
Assume as an axiom that a physical system always consists of
a natural number of $n$ quanta which are in a single pure state
among $q$ others.
Any single such particle is thus the carrier of
exactly one $q$-it, henceforth called ``quit.''
(In the spin-one half case, this reduces to the bit.)
That is, encoded in such a quantum system are always
$n$ quits of information.
The quit is an irreducible amount of classical and quantum information.
The quits need not be located at any single particle
(i.e., one quit per particle),
but they may be spread over the $n$ particles \cite{DonSvo01}.
In this case one calls the state of the particles ``entangled.''
According to Schr\"odinger's own
interpretation \cite{schrodinger},
the quantum wave function (or quantum state) is a
``catalogue of expectation values''
about this state; and in particular about the quits.
Since an experimenter's knowledge about a quantum system may be very
limited, the experimenter might not have operational access
to the ``true'' pure state out there.
(In particular, it need not be clear which questions have to be ask
to sort out the valid pure state from other ones.)
This ignorance on the experimenter's side is characterized by a nonpure
state.
Thus one should differentiate between the ``true'' quantum state out
there and the experimenter's ``poor man's version'' of it.
Both type of states undergo a unitary time evolution, but their
ontological status is different.

Why has the no-measurement interpretation of quantum mechanics
been not wider accepted and has attracted so little attention so far?
One can only speculate about the reasons.

For one thing, the interpretation seems to have no operational, testable
consequences.
Indeed, hardly any interpretation does.
So, what is any kind of interpretation of some formalism good for if
it cannot be operationalized?

Think of the Everett interpretation of quantum mechnics, which is
nevertheless highly appreciated among some circles, mainly in the
quantum computation community.
It has to offer no operationalizable consequences, just mindboggling
scenarios.

Or consider Bohr's ``Copenhagen'' interpretation, whatever that means to
its successors or to Bohr himself.
It is the canonical interpretation of  quantum mechanics,
a formalism
co-created by people, most notably  Einstein, Schr\"odiner and De
Broglie, who totally disagreed with that interpretation.
This does not seem to be the case for Heisenberg and von
Neumann. The latter genius  even
attempted to state an inapplicable theorem directed against hidden
parameters to support some of Bohr's tendencies.
Nowadays, many eminent researchers in the foundations of quantum
mechanics still stick with Bohm's interpretation or whatever sense
they have made out of it.
But does Bohr's ``Copenhagen'' interpretation have any operational
consequences?

With the advent of quantum information theory, the notion of information
seems to be the main interpretational concept.
Consequently, information interpretations of quantum mechanics begin to
be widespread.
Yet, despite the heavy use of the term ``information,''
the community does not seem to have
settled upon an unambiguous meaning of the term
``information.''
And also in this case, the interpretations do
not seem to have operational consequences.

Many recent developments in quantum information theory
are consistent
with the no-measurement interpretation.
Unitarity and the associated
one-to-onenness even for one
quantum events seems to be the guiding principle.
Take, for example, the no-cloning theorem, quantum teleportation,
entanglement swapping,  purification and so on
\cite{Gruska,Nielsen-book}.
Actually, the no-measurement interpretation seems to promote the
search for new phenomena in this regime, and might thus contribute to
a progressive research program.

Indeed, it is the author's belief  that being helpful in developing
novel theories and testing
phenomena is all one can ever hope for a good interpretation.
Any ``understanding'' of or ``giving meaning'' to
the formalism is desireable only
to the effect that it yields new predictions, phenomena
and technologies.
And in this sense,
the no-measurement interpretation claiming the
conventionality of quantum measaurements
should be perceived.
It too cannot offer direct operationalizable consequences,
yet may facilitate thoughts in new, prosperous directions.

\section{Summary}
We have reviewed conventions in two of the dominating theories of
contemporary physics, the theory of relativity and quantum mechanics.
In relativity theory we suggest to accept the constancy of one
particular speed as a convention.
Lorentz-type transformation laws can then be geometrically
derived under mild side assumptions.
In order for a generalied
principle of relativity and thus generalized form invariance to hold,
the particular signalling type entering the transformations
should correspond to the
dominating type of physical interactions.

The
no-measurement interpretation of quantum mechanics  suggests that there
is no such thing
as an irreversible measurement. In fact, there is no measurement at all,
never.
This kind of irreversibility
associated with the measurement
process is just an idealistic, subjective construction on the
experimenter's side to express the
for-all-practical-purposes impossibility to undo a measurement.

\section*{Appendix. Proof of Alexandrov's theorem}

Alexandrov's theorem states that,
for ${\Bbb R}^n$, $n\ge 3$ with the metric ${\rm diag}(+,+,+,\cdots ,+,-)$
and a one-to-one map $r\mapsto r'$ preserving light cones (i.e., zero distance)
such that  for all $r,s\in  {\Bbb R}^n$,
$$(r-s,r-s)=0 \Longleftrightarrow (r'-s',r'-s')=0,$$
$r\mapsto r'$ is essentially a Lorentz transformation;
i.e., it has the form $r\mapsto r'=\alpha Lr+a$ for some nonzero $\alpha \in {\Bbb R}$. $a\in {\Bbb R}^n$,
and a linear one-to-one map $L:{\Bbb R}^n\mapsto {\Bbb R}^n$ satisfying $(Lr,Ls)=(r,s)$ for all $r$
and $s$ in ${\Bbb R}^n$.

In what follows we shall review a complete proof of Alexandrov's theorem
very similar to the one sketched by Lester   \cite{lester}.
The proof consists of three stages:
\begin{itemize}
\item[(I)] a proof that, given ${\Bbb R}^n$, $n\ge 3$ with the metric ${\rm diag}(+,+,\cdots ,+,-)$
and a one-to-one map preserving light cones (i.e., zero distance),
all lines are mapped onto lines;

\item[(II)] a proof of the {\em fundamental theorem of affine geometry} stating that
a one-to-one map from ${\Bbb R}^n$, $n\ge 2$ onto itself which maps all lines onto lines must be {\em affine};
i.e., must be a {\em linear} map and a translation;

\item[(III)] a proof that, given ${\Bbb R}^n$, $n\ge 2$ with the metric ${\rm diag}(+,\cdots ,+,-)$
and a {\em linear} one-to-one map preserving a single light cone (i.e., zero distance)
must be essentially a Lorentz transformation (up to a translation and a dilatation);
i.e., it has the form $r\mapsto \alpha Lr+a$ for some nonzero $\alpha \in {\Bbb R}$. $a\in {\Bbb R}^n$,
and a linear one-to-one map $L:{\Bbb R}^n\mapsto {\Bbb R}^n$ satisfying $(Lr,Ls)=(r,s)$ for all $r$
and $s$ in ${\Bbb R}^n$.
\end{itemize}

In what follows, a constant translation is
taken account of by addition of a vector $a\in {\Bbb R}^n$.
The remaining transformation preserves the origin; i.e.,
$0\mapsto 0'$.
We shall often refer to this remaining transformation
(after the constant parallel shift moving the map of the origin back into the origin)
simply as (homogeneous) transformation.
(Note that if $f:r\mapsto \alpha L r+a$, then the homogeneous part
os obtained by subtracting $a=f(0)$.)
This  constant shift $a$ has to be added to the final mapping.


The geometric proof of (I) proceeds in five steps, covering the mapping of
(i) lightlike (null) lines onto lightlike lines;
(ii) lightlike (null) planes onto null planes;
(iii) spacelike lines onto spacelike lines;
(iv) timelike planes onto planes;
and finally (v) timelike lines onto lines.

In what follows, the configurations are demonstrated for ${\Bbb R}^3$
with the metric $(r,s)=r_1s_1+r_2s_2-(1/c^2)r_3s_3$.
For arbitrary dimensions we refer to \cite{benz}.
In this section, the velocity of light $c$ will be set to unity; i.e.,
$c=1$.
The terms ``null'' and ``lightlike'' will be used synonymously.

To show (i) let us first define a {\em null cone} with {\em vertex} $a$ by
$${\cal C}(a)=\left\{ r\in {\Bbb R}^3  \mid (r-a,r-a)=0\right\}.$$
By assumption, light cones are preserved, i.e., ${\cal C}(a)\leftrightarrow {\cal C}(a')$.
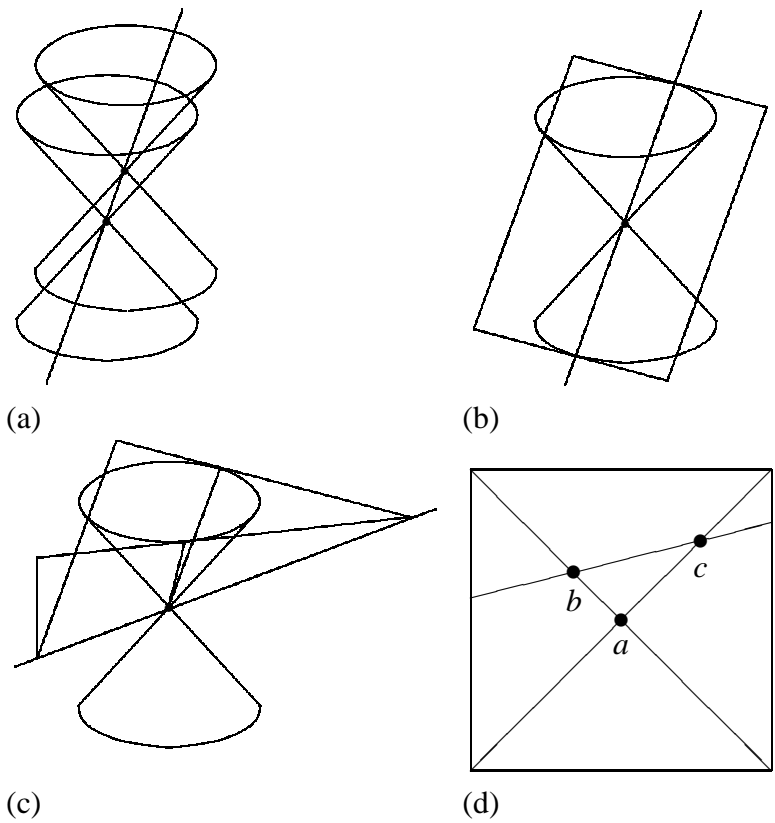
\begin{figure}
\begin{center}
\begin{tabular}{ll}
\unitlength 0.80mm
\linethickness{0.4pt}
\begin{picture}(38.44,61.94)
\bezier{88}(5.33,50.00)(-4.67,45.00)(5.33,40.00)
\bezier{84}(5.33,50.00)(15.33,53.33)(25.33,50.00)
\bezier{88}(25.33,50.00)(35.33,45.00)(25.33,40.00)
\bezier{84}(5.33,40.00)(15.33,36.67)(25.33,40.00)
\bezier{36}(0.35,11.03)(-0.06,7.89)(5.31,5.96)
\bezier{40}(5.31,5.96)(7.44,4.85)(15.24,4.14)
\multiput(0.27,11.02)(0.12,0.13){245}{\line(0,1){0.13}}
\bezier{36}(30.31,11.03)(30.71,7.89)(25.34,5.96)
\bezier{40}(25.34,5.96)(23.21,4.85)(15.41,4.14)
\multiput(30.53,11.02)(-0.12,0.13){246}{\line(0,1){0.13}}
\multiput(5.36,0.33)(0.12,0.33){189}{\line(0,1){0.33}}
\bezier{88}(8.44,58.30)(-1.56,53.30)(8.44,48.30)
\bezier{84}(8.44,58.30)(18.44,61.63)(28.44,58.30)
\bezier{88}(28.44,58.30)(38.44,53.30)(28.44,48.30)
\bezier{84}(8.44,48.30)(18.44,44.97)(28.44,48.30)
\bezier{36}(3.46,19.33)(3.05,16.19)(8.42,14.26)
\bezier{40}(8.42,14.26)(10.55,13.15)(18.35,12.44)
\multiput(3.38,19.32)(0.12,0.13){245}{\line(0,1){0.13}}
\bezier{36}(33.42,19.33)(33.82,16.19)(28.45,14.26)
\bezier{40}(28.45,14.26)(26.32,13.15)(18.52,12.44)
\multiput(33.64,19.32)(-0.12,0.13){246}{\line(0,1){0.13}}
\put(15.32,27.30){\circle*{1.21}}
\put(18.35,35.79){\circle*{1.21}}
\end{picture}
&
\unitlength 0.80mm
\linethickness{0.4pt}
\begin{picture}(49.27,61.61)
\bezier{88}(15.66,49.67)(5.66,44.67)(15.66,39.67)
\bezier{84}(15.66,49.67)(25.66,53.00)(35.66,49.67)
\bezier{88}(35.66,49.67)(45.66,44.67)(35.66,39.67)
\bezier{84}(15.66,39.67)(25.66,36.34)(35.66,39.67)
\bezier{36}(10.68,10.70)(10.27,7.56)(15.64,5.63)
\bezier{40}(15.64,5.63)(17.77,4.52)(25.57,3.81)
\multiput(10.60,10.69)(0.12,0.13){245}{\line(0,1){0.13}}
\bezier{36}(40.64,10.70)(41.04,7.56)(35.67,5.63)
\bezier{40}(35.67,5.63)(33.54,4.52)(25.74,3.81)
\multiput(40.86,10.69)(-0.12,0.13){246}{\line(0,1){0.13}}
\multiput(15.69,-0.00)(0.12,0.33){189}{\line(0,1){0.33}}
\put(25.65,26.97){\circle*{1.21}}
\multiput(17.00,54.82)(0.45,-0.12){71}{\line(1,0){0.45}}
\multiput(0.48,9.37)(0.45,-0.12){72}{\line(1,0){0.45}}
\multiput(32.75,0.88)(0.12,0.33){138}{\line(0,1){0.33}}
\multiput(0.48,9.22)(0.12,0.33){138}{\line(0,1){0.33}}
\end{picture}
\\
(a)&(b)\\
\unitlength 0.80mm
\linethickness{0.4pt}
\begin{picture}(69.24,54.85)
\bezier{88}(15.66,49.67)(5.66,44.67)(15.66,39.67)
\bezier{84}(15.66,49.67)(25.66,53.00)(35.66,49.67)
\bezier{88}(35.66,49.67)(45.66,44.67)(35.66,39.67)
\bezier{84}(15.66,39.67)(25.66,36.34)(35.66,39.67)
\bezier{36}(10.68,10.70)(10.27,7.56)(15.64,5.63)
\bezier{40}(15.64,5.63)(17.77,4.52)(25.57,3.81)
\multiput(10.60,10.69)(0.12,0.13){245}{\line(0,1){0.13}}
\bezier{36}(40.64,10.70)(41.04,7.56)(35.67,5.63)
\bezier{40}(35.67,5.63)(33.54,4.52)(25.74,3.81)
\multiput(40.86,10.69)(-0.12,0.13){246}{\line(0,1){0.13}}
\put(25.65,26.97){\circle*{1.21}}
\multiput(0.00,17.27)(0.32,0.12){219}{\line(1,0){0.32}}
\multiput(16.82,54.85)(0.46,-0.12){107}{\line(1,0){0.46}}
\multiput(16.82,54.70)(-0.12,-0.33){108}{\line(0,-1){0.33}}
\multiput(25.61,27.12)(0.12,0.32){72}{\line(0,1){0.32}}
\multiput(25.61,26.97)(0.12,0.48){23}{\line(0,1){0.48}}
\multiput(65.30,42.00)(-1.10,-0.12){56}{\line(-1,0){1.10}}
\put(3.74,35.30){\line(0,-1){16.52}}
\end{picture}
&
\unitlength 0.8mm
\linethickness{0.4pt}
\begin{picture}(50.00,50.00)
\put(0.00,0.00){\line(1,0){50.00}}
\put(0.00,0.00){\line(0,1){50.00}}
\put(50.00,0.00){\line(0,1){50.00}}
\put(50.00,50.00){\line(-1,0){50.00}}
\put(0.00,50.00){\line(1,-1){50.00}}
\put(50.00,50.00){\line(-1,-1){50.00}}
\put(0.00,28.67){\line(4,1){50.00}}
\put(17.07,32.93){\circle*{2.00}}
\put(38.17,38.17){\circle*{2.00}}
\put(25.00,25.00){\circle*{2.00}}
\put(25.00,20.50){\makebox(0,0)[cc]{$a$}}
\put(17.17,28.33){\makebox(0,0)[cc]{$b$}}
\put(38.17,33.33){\makebox(0,0)[cc]{$c$}}
\end{picture}
\\
(c)&(d)
\end{tabular}
\end{center}
\caption{\label{fig1-2001-conven}
Illustrations of the proof that
(a) lightlike (null) lines into lightlike lines; (b) lightlike (null) planes into null planes;
(c) spacelike lines into spacelike lines;
(d) timelike planes map into planes.}
\end{figure}

As illustrated in Fig. \ref{fig1-2001-conven}(a),
any null (lightlike) line is the intersection of two tangent null cones.
Since null cones are preserved, so are null (lightlike) lines.
Thus, null lines are mapped into null lines.
The same is true for the inverse map.
Hence, null lines are mapped {\em onto} null lines.
(The same is true for the other proof steps as well
but will not be mentioned explicitly.)

To show (ii), notice that, as illustrated in Fig. \ref{fig1-2001-conven}(b),
a null cone with vertex on some null plane
is tangent to that plane along a null line.
Points of ${\Bbb R}^3$ are on the null plane
if and only if they either lie on this null line
or on no null cone with vertex on this line.
The latter sentence could be understood as follows.
Imagine any point  of ${\Bbb R}^3$ outside of the null plane
(either ``below'' or ``above'').
Any such point is element of some null cone with vertex
on the null line mentioned.
On the contrary, any point on the null plane cannot be reached by such null cones
(except the ones located on the null line mentioned),
but by other null cones whose vertex is not on that null line.
Null lines and cones are preserved; thus null planes are preseved as well.

To show (iii), notice that, as illustrated in Fig. \ref{fig1-2001-conven}(c),
any spacelike line is the intersection of two null planes.
Since null planes are preserved, spacelike lines are preserved.

To show (iv), notice that, as illustrated in Fig. \ref{fig1-2001-conven}(d),
the points in a timelike plane are covered by infinitely many intersecting null and spacelike
lines in that plane.
By fixing, for instance, a triangle formed by the vertices $a,b,c$
of three such lines
(e.g., two null and one spacelike line) ``spans'' the timelight plane.
Because of the one-to-oneness of the mapping,
the image of the triangle with the vertices $a',b',c'$
``spans'' the transformed plane (different points are mapped onto different points).
Therefore, the three lines  forming the transformed
triangle must be coplanar.
In general, the images of all lines lying in the original timelight plane must be coplanar.
Thus, timelike planes map into planes.

To show (v), notice that any timelight line is the intersection of two timelight planes.
Since timelike planes are mapped onto planes, they intersect into a line.
Thus, any timelike line is mapped into a line.

In summary, all three types of lines---lightlike (null), spacelike
and timelike lines---are mapped onto lines.
(Recall that the same arguments apply for the inverse transformation as well.)


The geometric proof of (II), in particular
the linearity of the transformation
proceeds
from the preservation of lines
essentially by utilizing the preservation of {\em parallelism}
among lines.
As will be demonstrated below, the preservation of parallelism
implies that the transformation is additive.
The associated transformation of the field ${\Bbb R}$
is an automorphism.
It then only remains to be proven that the only automorphism of ${\Bbb R}$
is the identity function.

Let us first introduce some notation.
For a much more comprehensive approach the reader is refered to the literature
(e.g., the book by Gruenberg \& Weir \cite{Gruenberg-77}).
Let $a$ be a fixed ``translation'' vector of ${\Bbb R}^n$
and $M$ be a linear subspace of ${\Bbb R}^n$.
[Recall that a subset $S\subset {\Bbb R}^n$ is called a (linear)
subspace if $S$ is a vector space in its own right with respect to the
same vector addition and scalar multiplication than ${\Bbb R}^n$.]
Then $a+M$ denotes the set of all vectors
$a+M= \left\{ a+m \mid m\in M\right\}$.
It is called  {\em translated subspace} or {\em coset} or {\em affine subspace}
of ${\Bbb R}^n$.
The dimension of a translated subspace $a+M$ is the dimension of the
linear subspace $M$; i.e., ${\rm dim}(a+M)={\rm dim}(M)$.
Translated subspaces of dimensions $0,1,2$ are called
{\em points, lines} and {\em planes}, respectively.
Let the {\em join} $S_1\circ S_2$ of two  translated subspaces
$S_1,S_2$ be the intersection
of all translated subspaces in ${\Bbb R}^n$ which contain both $S_1$ and
$S_2$.
(The join is again a translated subspace.)
Furthermore,
if $S\subset {\Bbb R}^n$ is any set of vectors in ${\Bbb R}^n$, we
denote by the (linear) {\em span} ${\rm span}(S)$
the intersection of all the subspaces of ${\Bbb R}^n$ which contain $S$.


We shall call an {\em automorphism}
a one-to-one mapping of ${\Bbb R}^n$ {\em onto itself} preserving all translated
subspaces.
The {\em fundamental theorem of affine geometry}
(e.g., ref. \cite[Theorem 5]{Gruenberg-77}) states that, for
${\Bbb R}^n$,
$n\ge 2$, any automorphism
 induces a
{\em linear} transformation $L$
and a translation vector $a$ such that
$r\mapsto r'=\alpha Lr+a$.

In what follows, a proof of the fundamental theorem of affine geometry
will be given for the case of the vector space ${\Bbb R}^n$, $n\ge 2$
with field ${\Bbb R}$.
First, a proof will be given that any such automorphism
of ${\Bbb R}^n$ implies an automorphism on the field of reals ${\Bbb R}$
(a definition will be given below).
By invoking the preservation of parallelism
one obtains both the uniqueness of the associated mapping of the
field ${\Bbb R}$ onto itself and furthermore
the additivity of the transformation as a whole.

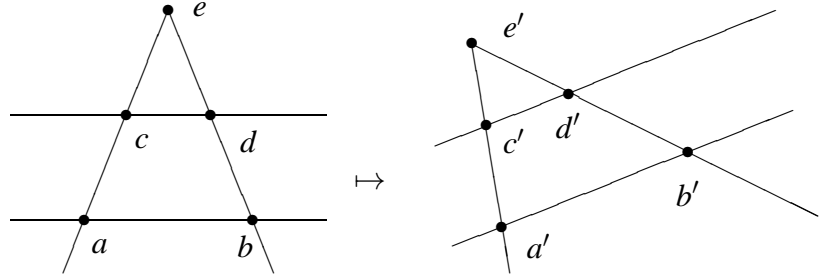
\begin{figure}
\begin{center}
\unitlength 0.7mm
\linethickness{0.4pt}
\begin{picture}(153.33,51.00)
\put(0.00,10.00){\line(1,0){60.00}}
\put(0.00,30.00){\line(1,0){60.00}}
\put(10.00,0.00){\line(2,5){20.00}}
\put(30.00,50.00){\line(2,-5){20.00}}
\put(14.00,10.00){\circle*{2.00}}
\put(46.00,10.00){\circle*{2.00}}
\put(38.00,30.00){\circle*{2.00}}
\put(30.00,50.00){\circle*{2.00}}
\put(22.00,30.00){\circle*{2.00}}
\put(17.00,5.00){\makebox(0,0)[cc]{$a$}}
\put(44.67,5.00){\makebox(0,0)[cc]{$b$}}
\put(25.00,25.00){\makebox(0,0)[cc]{$c$}}
\put(45.33,25.00){\makebox(0,0)[cc]{$d$}}
\put(36.00,50.00){\makebox(0,0)[cc]{$e$}}
\put(84.00,5.00){\line(5,2){64.67}}
\put(80.67,24.00){\line(5,2){64.67}}
\put(87.67,43.33){\line(1,-6){7.22}}
\put(88.00,43.33){\line(2,-1){65.33}}
\put(87.67,43.67){\circle*{2.00}}
\put(90.33,28.00){\circle*{2.00}}
\put(93.33,8.67){\circle*{2.00}}
\put(128.67,23.00){\circle*{2.00}}
\put(106.00,34.00){\circle*{2.00}}
\put(95.67,47.00){\makebox(0,0)[cc]{$e'$}}
\put(100.33,5.00){\makebox(0,0)[cc]{$a'$}}
\put(128.67,15.00){\makebox(0,0)[cc]{$b'$}}
\put(95.67,25.00){\makebox(0,0)[cc]{$c'$}}
\put(106.00,28.00){\makebox(0,0)[cc]{$d'$}}
\put(68.33,17.00){\makebox(0,0)[cc]{$\mapsto$}}
\end{picture}
\end{center}
\caption{\label{fig1a-2001-conven}
Geometrical proof of the preservation of parallelism due to the
preservation of lines.}
\end{figure}

Note that the automorphism preserves parallellism.
This can be seen by ``fixing'' appropriate four points $a,b,c,d$ on two
lines
which are originally parallel, drawing two nonparallel lines through
them which meet in another point $e$.
Since by assumption all lines are preserved, so are their meeting points
$a',b',c',d'$.
Furthermore, because of bijectivity,
two parallel lines have no point in common.
Thus, the two lines which are originally parallel are
mapped onto colpanar lines which are disjoint; i.e., they are again parallel.
Hence, parallelism is conserved.
A concrete configuration
illustrating this geometrical argument is drawn in Fig. \ref{fig1a-2001-conven}.

Consider an arbitrary nonzero vector $a\in {\Bbb R}^n$.
According to the assumptions, any line
$0\circ a= {\rm span}(a)$
is transformed into a line
$0'\circ a'= {\rm span}(a')$,
thereby inducing a one-to-one mapping of
all points of ${\rm span}(a)$
onto the points of ${\rm span}(a')$.
That is, the transformation defines a one-to-one mapping
\begin{equation}
\zeta :x\mapsto x'
\end{equation}
of the field of real numbers onto itself by the definition
\begin{equation}
(xa)'=x'a'.  \label{2001-convention-amb}
\end{equation}
It immediately follows that
$\zeta :0\mapsto 0'$
as well as
$\zeta :1\mapsto 1'$.
It will be shown that $\zeta$ is an automorphism; i.e., a one-to-one mapping
of ${\Bbb R}$ onto itself with the properties that
$\zeta (x+y)= \zeta (x) + \zeta (y)$, as well as
$\zeta (xy)= \zeta (x)  \zeta (y)$.

\begin{figure}
\begin{center}
\unitlength 0.50mm
\linethickness{0.4pt}
\begin{picture}(100.00,75.67)
\put(10.00,10.00){\vector(1,0){50.00}}
\put(60.00,10.00){\vector(1,0){40.00}}
\put(60.00,10.00){\line(-1,2){16.67}}
\put(10.00,10.00){\vector(1,1){33.33}}
\put(43.33,43.33){\vector(1,1){26.67}}
\put(100.00,10.00){\line(-1,2){30.00}}
\put(60.00,5.00){\makebox(0,0)[cc]{$a$}}
\put(41.33,50.00){\makebox(0,0)[cc]{$b$}}
\put(66.67,75.67){\makebox(0,0)[cc]{$xb$}}
\put(100.00,5.00){\makebox(0,0)[cc]{$xa$}}
\put(10.00,5.00){\makebox(0,0)[cc]{$0$}}
\put(58.00,27.00){\makebox(0,0)[lc]{$a\circ b$}}
\put(91.67,47.00){\makebox(0,0)[lc]{$xa\circ xb$}}
\end{picture}
\end{center}
\caption{\label{fig1aa-2001-conven}
Rescaling does not effect parallelism.}
\end{figure}
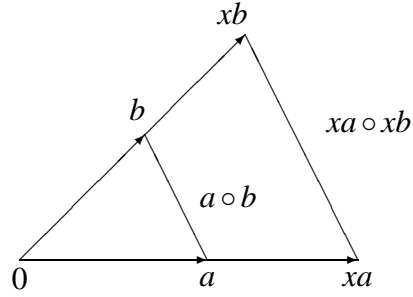

First it
is shown that $\zeta$ does not depend on the particular choice of $a\in
{\Bbb R}^n$.
(i) Case 1:
Consider two linearly independent vectors $a,b$ of ${\Bbb R}^n$,
$(xa)'=x'a'$
and
$(xb)'=x''b'$, $x'\neq x''$.
Since $0=0'=0''$, one can assume that $x\neq 0$.
The join $xa \circ xb$ is the intersection  of all the
subspaces of ${\Bbb R}^n$ containing both $xa$ and $xb$.
Since $xa$ and $xb$ are vectors, this is just the subspace
spanned by the line joining them.
$xa \circ xb$ is parallel to $a \circ b$.
(Rescaling does not affect parallelism; cf. Fig. \ref{fig1aa-2001-conven}.)
The transformation preserves parallelism, and therefore
$a' \circ b'$ must also be parallel to
$x'a \circ x'b'$ and
$x'a \circ x''b'$, the lines connecting
$x'a $ with $ x'b'$
and $x'a $ with $ x''b'$.
This can only be satisfied for $x'=x''$.
Hence, $\zeta$ is independent
of the argument and only depends on the transformation.

(ii) Case 2:
Consider two linearly dependent vectors $a,b$ of ${\Bbb R}^n$.
In this case, choose a third vector $c$ which does not lie in the linear
subspace ${\rm span}(a)$ spanned by $a$ and $b$.
Then, by the argument used in case 1, $\zeta$ is the same for
$a,c$ and $b,c$; thus $\zeta$ is also the same for $a,b$.
Hence, to sum up the finding in the two cases, $\zeta$ is independent
of the argument vector and only depends on the transformation.

\begin{figure}
\begin{center}
\unitlength 0.7mm
\linethickness{0.4pt}
\begin{picture}(130.00,44.67)
\put(0.00,0.00){\line(2,3){20.00}}
\put(20.00,30.00){\line(5,1){25.00}}
\put(0.00,0.00){\line(5,1){25.00}}
\put(25.00,5.00){\line(2,3){20.00}}
\put(80.00,0.00){\line(5,1){50.00}}
\put(130.00,10.00){\line(-2,3){20.00}}
\put(80.00,0.00){\line(-2,3){20.00}}
\put(60.00,30.00){\line(5,1){50.00}}
\put(0.00,6.67){\makebox(0,0)[cc]{$0$}}
\put(27.33,2.00){\makebox(0,0)[cc]{$a$}}
\put(19.00,34.00){\makebox(0,0)[cc]{$b$}}
\put(47.33,39.33){\makebox(0,0)[cc]{$a+b$}}
\put(49.67,15.67){\makebox(0,0)[cc]{$\mapsto$}}
\put(72.33,0.00){\makebox(0,0)[cc]{$0'$}}
\put(60.67,33.67){\makebox(0,0)[cc]{$b'$}}
\put(111.33,44.67){\makebox(0,0)[cc]{$a'+b'$}}
\put(130.00,3.67){\makebox(0,0)[cc]{$a'$}}
\end{picture}
\end{center}
\caption{\label{fig1b-2001-conven}
Geometrical proof of the preservation of any parallelogram and of
additivity due to the conservation of parallelism.}
\end{figure}
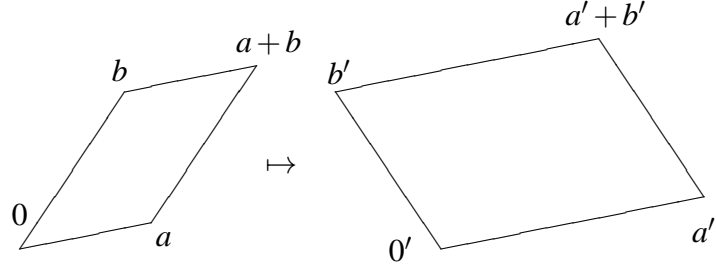

We shall pursue the proof that, given the preservation of lines,
the associated mapping is additive (up to translations).
A geometric interpretation
of this proof is drawn in Fig. \ref{fig1b-2001-conven}.
(i) Case 1:
If $a$ and $b$ are
linearly independent nonzero vectors (the zero vector case is trivial)
of
${\Bbb R}^n$,
then the parallelogram $a,0,b,a+b$ is mapped into
the parallelogram $a',0',b',a'+b'$
and
\begin{equation}
(a+b)'=a'+b'.\label{2001-convention-aub}
\end{equation}
This is also true if $a$ or $b$ is the zero vector.

(ii) Case 2:
If $a$ and $b$ are linearly dependent and nonzero,
choose a third vector $c\not\in {\rm span}(a)$
(so that $c$ is linearly independent of $a$ and $b$),
and apply the above considerations for the pairs
$a+b$ \& $c$ rendering $a+b+c\mapsto (a+b)'+c'$,
$a$ \& $b+c$ rendering $a+b+c\mapsto a'+(b+c)'$,
$b$ \& $c$ rendering $b+c\mapsto b'+c'$,
such that $(a+b)'+c'= a'+b'+c'$, which is satisfied only if
again $(a+b)'= a'+b'$.

Two further properties assuring that $\zeta$ is an automorphism
 can be deduced from the uniqueness of
Eq. (\ref{2001-convention-amb})
and Eq. (\ref{2001-convention-aub}) and the usual axioms of linear
vector spaces.
(i) Automorphism property 1:
Let $a'\neq 0$, then for all $x,y\in {\Bbb R}$,
\begin{equation}
(x+y)'a'=[(x+y)a]'=(xa+ya)'=(xa)'+(ya)'=x'a'+y'a'=(x'+y')a'
\end{equation}
and thus
\begin{equation}
(x+y)'=x'+y'.
\end{equation}
(ii) Automorphism property 2:
By the assumption of vector spaces,
$(xy)a=x(ya)$  for all $x,y\in {\Bbb R}$ and $a\in {\Bbb R}^n$.
Therefore,
\begin{equation}
(xy)'a'=x'(ya)'=x'(y'a')=(x'y')a'
\end{equation}
and thus
\begin{equation}
(xy)'=x'y'.
\end{equation}

In order to complete the proof of linearity, it will be shown that the only
automorphism of the field ${\Bbb R}$ into itself is the identity function ${\rm id}:x\mapsto
x$.
This can be demonstrated by realizing that
the algebraic properties of neutral
elements $0,1$ with regard to addition and multiplication have to be
preserved; i.e.,
$0\mapsto 0$
and
$1\mapsto 1$.
Furthermore, since $1$ has to be preserved and any natural number $n\in {\Bbb N}$
is the sum of $n$ $1$'s, ${\Bbb N}$ has only a single automorphism--the identity
function
${\rm id}$.
A very similar argument  holds for ${\Bbb Z}$.
Since any element of the positive rationals
can be represented by the quotient $n/m$ with $n,m\in {\Bbb N}$,
again ${\Bbb Q}$ has only a single automorphism--the identity
function ${\rm id}$.
In order to be able to obtain the same result for  ${\Bbb R}$,
one has to make sure the Dedekind construction of
the reals works; in particular the
preservation of Dedekind sections.
This requires the preservation of the order relation ``$<$'' in ${\Bbb R}$,
which is equivalent to the preservation of positivity, since
$x< y$ can always be rewritten into $0< y-x$.
Notice that every
positive $0< x\in {\Bbb R}$ can be written as $x=y^2$, $y\in {\Bbb R}$
$y\neq 0$.
Since $x=y^2$ is mapped onto $x'=(y^2)'=(y')^2$ with $y'\neq 0$ (recall
that $0\mapsto 0$), $x'>0$.
This allows the Dedekind construction of the reals using the
rationals, which in turn yields the desired fact that
${\Bbb R}$ has only a single automorphism--the identity
function ${\rm id}$.
(This is not true for example for ${\Bbb C}$,
since for example $x+iy\mapsto x-iy$ is an automorphism but not the
identity.)

A way to get rid of the factor $\alpha$ is by considering the tangent hyperboloid
$x^2+y^2-z^2=1$ of the null cone $x^2+y^2-z^2=0$, translating it once and then back
to the original figure. The requirement that this should result in the same hyperboloid
fixes $\alpha$
\cite{havlicek-priv}.


We shall now concentrate on a proof of (III).
Let us first note that, in the case of a linear map, the preservation of a {\em single}
light cone is a sufficient condition for the preservation of all of them.
For, given the transformation
$x\mapsto \alpha Lx +a$, any shift of the null cone ${\cal C}(p)$ with vertex $p$
by a vector $s=q-p$ results in a null cone ${\cal C}(q)={\cal C}(p)+s$ with vertex $q=p+s$.
The latter null cone ${\cal C}(q)$ is mapped onto the null cone
\begin{eqnarray}
{\cal C}(q)\mapsto \alpha L{\cal C}(q) +a
&=& \alpha L ({\cal C}(p)+s)+a\nonumber \\
&=& \alpha L {\cal C}(p)+\alpha L s+a\nonumber \\
&=& (\alpha L {\cal C}(p)+a)+\alpha L s\nonumber \\
&=&  {\cal C}(p')+\alpha L s,\nonumber
\end{eqnarray}
which again is a null cone.

Recall that in Einstein's original work \cite[par 3]{ein-05},
linearity was never derived but was assumed for physical reasons.
{\em ``In the first place it is clear that the equations must
be {\em linear} on account of the properties of homogeneity which we attribute to space and time.''
[[``Zun\"achst ist klar, da\ss$\;$ die Gleichungen {\em linear} sein m\"ussen
wegen der Homogenit\"atseigenschaften, welche wir Raum und Zeit beilegen.'']]}
In what follows we shall closely follow Einstein's original argument
rendering $L$ to be the Lorentz transformations.

Take the standard four dimensional space-time case ${\Bbb R}^4$, and
consider, for the sake of simplicity,  the
quasi-twodimensional case (one space and the time coordinate)
of the constant motion along the $x$-axis of $K$
with velocity $v$ of a coordinate frame
$K'$ with the components $(x',y',z',t')$
against another coordinate frame ``at rest''
$K$ with the components $(x,y,z,t)$.
(Otherwise, $K$ can be
rotated such that the direction of motion is along the $x$ axis.)
Again, $c$ stands for the velocity of light.

Now define a particular series (in time) of points $\bar{x}=x-vt$.
Notice that the ``worldlines'' $(x=vt,0,0,t)$ just mark the
parametrization by the time parameter $t$
of all points at rest with respect to the moving frame $K'$.
That is, any such point has constant $\bar{x},y,z$ throughout all times
$t$.
It is sometimes convenient (cf. below) to
write the parameters of events in terms of
$(\bar{x},y,z,t)$
instead of
$(x,y,z,t)$.

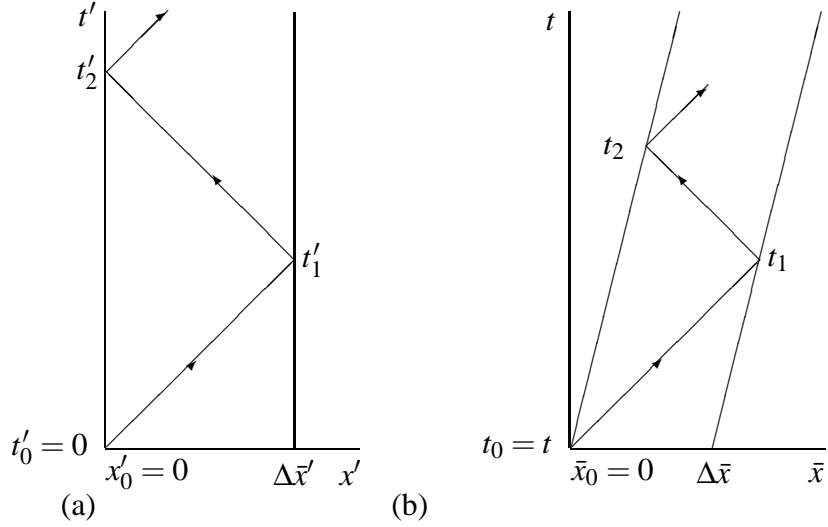
\begin{figure}
\begin{center}
\begin{tabular}{ll}
\unitlength 0.50mm
\linethickness{0.4pt}
\begin{picture}(77.34,123.00)
\put(10.00,6.67){\line(1,1){50.00}}
\put(60.00,56.67){\line(-1,1){50.00}}
\put(10.00,106.67){\line(1,1){16.33}}
\put(9.67,122.67){\line(0,-1){116.00}}
\put(60.00,6.67){\line(0,1){115.67}}
\put(32.67,28.37){\vector(1,1){1.33}}
\put(39.67,77.17){\vector(-3,4){1.50}}
\put(25.00,121.87){\vector(3,2){1.00}}
\put(5.00,6.67){\makebox(0,0)[rc]{$t'_0=0$}}
\put(65.00,56.67){\makebox(0,0)[cc]{$t'_1$}}
\put(5.00,106.67){\makebox(0,0)[cc]{$t'_2$}}
\put(5.00,122.00){\makebox(0,0)[cc]{$t'$}}
\put(9.67,6.34){\line(1,0){67.67}}
\put(75.00,0.00){\makebox(0,0)[cc]{$x'$}}
\put(60.00,0.00){\makebox(0,0)[cc]{$\Delta \bar{x}'$}}
\put(10.00,0.00){\makebox(0,0)[cl]{$x'_0=0$}}
\end{picture}
& $
\qquad
\qquad
$
\unitlength 0.50mm
\linethickness{0.4pt}
\begin{picture}(77.34,122.67)
\put(10.00,6.67){\line(1,1){50.00}}
\put(9.37,122.67){\line(0,-1){116.00}}
\put(32.67,28.87){\vector(1,1){1.33}}
\put(39.67,77.17){\vector(-3,4){1.50}}
\put(5.00,6.67){\makebox(0,0)[rc]{$t_0=t$}}
\put(65.00,56.67){\makebox(0,0)[cc]{$t_1$}}
\put(21.33,86.34){\makebox(0,0)[cc]{$t_2$}}
\put(9.67,6.34){\line(1,0){67.67}}
\put(75.00,0.00){\makebox(0,0)[cc]{$\bar{x}$}}
\put(47.67,0.00){\makebox(0,0)[cc]{$\Delta \bar{x}$}}
\put(10.00,0.00){\makebox(0,0)[cl]{$\bar{x}_0=0$}}
\put(9.67,6.33){\line(1,4){29.08}}
\put(47.30,6.33){\line(1,4){29.08}}
\put(60.00,56.67){\line(-1,1){30.33}}
\put(30.00,87.00){\line(1,1){16.33}}
\put(45.00,101.50){\vector(3,2){1.00}}
\put(4.67,120.00){\makebox(0,0)[cc]{$t$}}
\end{picture}
\\
(a)&(b)
\end{tabular}
\end{center}
\caption{\label{fig2-2001-conven}
Generation of radar coordinates by a light clock following Einstein's
procedures and conventions (a) from within the system $K'$;
(b) the same procedure  seen from the system $K$.}
\end{figure}

Let us construct ``radar coordinates'' of $K'$ by utilizing a light
clock starting at some arbitrary point $\bar{x}=0$
at $t'_0$, traveling some distance $\Delta \bar{x}$ to a mirror, where
it arrives and is instantly reflected at $K'$-time $t'_1$ towards the
original source mirror and arrives there at  $K'$-time $t'_2$
[cf. Fig. \ref{fig2-2001-conven}(a)].
If one adopts the usual conventions for synchronization,
$t'_1$ is just the arithmetic mean of the two times
$t'_0$
and
$t'_2$; i.e.,
\begin{equation}
t_1={1\over 2}(t'_0+t'_2).
\label{e1-2001-conven}
\end{equation}

In order to find the transformation mapping $K$ onto $K'$,
rewrite the transformed coordinates as functions of the original system; e.g.,
$t'=t'(\bar{x},y,z,t)$.
In this parametrization, the coordinates are given by
\begin{eqnarray}
t'_0 &=& t' (0,0,0,t),
\label{e2-2001-conven}
\\
t'_1 &=& t' \left( \Delta\bar{x},0,0,t+{\Delta \bar{x} \over c^2-v^2}\right ),
\label{e3-2001-conven}
\\
t'_2 &=& t' \left( 0,0,0,t+{\Delta \bar{x} \over c^2-v^2}+{\Delta \bar{x} \over c^2+v^2}\right ),
\label{e4-2001-conven}
\end{eqnarray}
where
\begin{eqnarray}
t_0 &=& t, \nonumber \\
t_1 &=& t+{\Delta \bar{x} \over c^2-v^2} ,\nonumber \\
t_2 &=& t+{\Delta \bar{x} \over c^2-v^2}+{\Delta \bar{x} \over c^2+v^2} \nonumber
\end{eqnarray}
results from the following consideration.
The $K$-time $\Delta t_1 =t_1-t_0$ it takes for light to arrive at the
first mirror
is given by the total distance it takes for light to travel to it,
divided by the velocity of light.
Since the mirror travels with velocity
$v$,
\begin{equation}
\Delta t_1 ={\Delta \bar{x} +v \Delta t_1\over c}={\Delta \bar{x}\over c-v}.
\label{e4a-2001-conven}
\end{equation}
A similar argument yields
$$
\Delta t_1 = t_2-t_1=
{\Delta \bar{x} -v \Delta t_1\over c}={\Delta \bar{x}\over c+v}.
$$

Inserting Eqs.
(\ref{e2-2001-conven})--(\ref{e4-2001-conven})
into
(\ref{e1-2001-conven}) yields
\begin{equation}
t' \left( \Delta\bar{x},0,0,t+{\Delta \bar{x} \over c^2-v^2}\right )=
{1\over 2}\left[
t' (0,0,0,t)+
t' \left( 0,0,0,t+{\Delta \bar{x} \over c^2-v^2}+{\Delta \bar{x} \over c^2+v^2}\right )
\right].
\label{e5-2001-conven}
\end{equation}
$\Delta \bar{x}$ can be arbitrarily small.
A partial derivation of (\ref{e5-2001-conven}) by ${\partial t'\over \partial \Delta\bar{x}}$
in the limit of infinitesimal $\Delta \bar{x}$ yields
\begin{equation}
{\partial t'\over \partial \bar{x}}
+{1\over c-v}{\partial t'\over \partial t} =
{1\over 2}\left(
{1\over c-v}+{1\over c+v}
\right){\partial t' \over \partial t} ,
\label{e6-2001-conven}
\end{equation}
and thus
\begin{equation}
{\partial t'\over \partial \bar{x}}
+{1\over c^2-v^2}{\partial t'\over \partial t} = 0.
\label{e7-2001-conven}
\end{equation}
Likewise,
$
(\partial t'/ \partial y)
=
(\partial t'/ \partial z)
=0.
$
As a result of this and Eq. (\ref{e7-2001-conven}),
$t'$ must be a linear function of
$t$ and $\bar{x}$ of the form
\begin{equation}
t' ( \bar{x},y,z,t)= \alpha (v) \left(t-
{v\over c^2-v^2}\,\bar{x} \right) .
\label{e8-2001-conven}
\end{equation}
$\alpha (v)$ is a yet arbitrary scale factor depending only on $v$.
Note that, without loss of generality, the origins of $K$
and $K'$ has been chosen such that
$t=t'=0$.
By substituting the explicit parameters for $\bar{x}=x-vt$ one obtains
\begin{equation}
t' ( \bar{x},y,z,t)= \alpha (v)
{1 \over 1-{v^2\over c^2}}
\left(t-
{v\over c^2}\,x \right) .
\label{e9-2001-conven}
\end{equation}

The transformation rules of the $x'$ parameter can be obtained by
considering the propagation of a light ray in $K'$ which starts at the
origin of $K$ and $K'$ (same origins) and moves along the $x$- and
$x'$-axes. The convention of the constancy of the speed of light
requires
\begin{equation}
x'=ct' = \alpha (v) c\left(t-
{v\over c^2-v^2}\,\bar{x} \right).
\label{e10-2001-conven}
\end{equation}
Now recall that, in terms of the $K$-parameters, this propagation
of this light ray is given by Eq. (\ref{e4a-2001-conven}); i.e., by
$t=\bar{x}/(c-v)$ (the differences $\Delta$ can be omitted because of
the ray starting at the coordinate origins).
By substituting $t$ in (\ref{e10-2001-conven}) one obtains
\begin{equation}
x'
=
\alpha (v)
{c^2\over c^2-v^2}\,\bar{x}
=
\alpha (v)
{1 \over 1-{v^2\over c^2}}
\,\bar{x}
=
\alpha (v)
{1 \over 1-{v^2\over c^2}}
\left(x-vt\right)
.
\label{e11-2001-conven}
\end{equation}

Let us now turn to the transformation of coordinates $y,z$ perpendicular
to the direction of motion $x$.
Consider a light ray propagating along the $y'$-axis,
and hence $\bar{x}=0$.
\begin{figure}
\begin{center}
\unitlength 0.5mm
\linethickness{0.4pt}
\begin{picture}(30.00,50.00)
\put(10.00,10.00){\vector(1,2){20.00}}
\put(10.00,10.00){\vector(1,0){20.00}}
\put(10.00,10.00){\vector(0,1){40.00}}
\put(3.00,50.00){\makebox(0,0)[cc]{$v_y$}}
\put(28.00,35.00){\makebox(0,0)[cc]{$c$}}
\put(30.00,5.00){\makebox(0,0)[cb]{$v$}}
\end{picture}
\end{center}
\caption{\label{fig3-2001-conven}
Velocity $v_y$ of a light ray propagating along the positive $y'$ axis
of a system traveling with velocity $v$ along the $x$- and $x'$-axes.}
\end{figure}
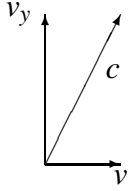
Inside the system $K$, the $y$-component of the light propagation
follows from the Pythagorean theorem, which is illustrated in
Fig. \ref{fig3-2001-conven}; i.e.,
$v_y=\sqrt{c^2-v^2}$.
Hence,
\begin{equation}
y'=ct'=\alpha (v) c\left(t-
{v\over c^2-v^2}\,\bar{x} \right),
\label{e12-2001-conven}
\end{equation}
for $\bar{x}=0$ and $t=y/v_y=y/\sqrt{c^2-v^2}$
\begin{equation}
y'
=
\alpha (v) {1\over \sqrt{1-{v^2\over c^2}}} \; y
.
\label{e12a-2001-conven}
\end{equation}
The same consideration applies to the transformation of the $z$- and
$z'$-axes.
Summing up, we obtain a transformation of the coordinates
$x\mapsto x'=Lx$ given by
\begin{equation}
L(v)=
\alpha (v) {1\over  \sqrt{1-{v^2\over c^2}}}
\left(
\begin{array}{cccc}
{1\over  \sqrt{1-{v^2\over c^2}}}&0&0&-{v\over  \sqrt{1-{v^2\over c^2}}}\\
0&1&0&0\\
0&0&1&0\\
-{v\over c^2 \sqrt{1-{v^2\over c^2}}}&0&0&{1\over  \sqrt{1-{v^2\over
c^2}}}
\end{array}
\right)
\label{e13-2001-conven}
\end{equation}

We now fix the factor $\alpha (v)$ by the conventional requirement that
a back-transformation should recover the original coordinates.
For this purpose we invent a third coordinate frame $K''$ which
propagates with the reverse (relative to $K'$) velocity $-v$ (measured
in $K$) along the $x$-, $x'$-, and its $x''$-axes.
The successive application of the transformation
(\ref{e13-2001-conven}) with
$L(v)$ and $L(-v)$ should bring back the coordinates to their original
form; i.e.,
\begin{equation}
L(v)L(-v)={\Bbb I}_4,
\label{e14-2001-conven}
\end{equation}
where ${\Bbb I}_4={\rm diag}(1,1,1,1)$ stands for the four-dimensional
unit matrix.
After evaluating the matrix product and comparing the coefficients,
one obtains
\begin{equation}
\alpha (v)\alpha (-v) = 1-{v^2\over c^2}.
\label{e15-2001-conven}
\end{equation}
That
$\alpha (v)=\alpha (\vert v\vert )$
only depends on the absolute value of the velocity can be seen by
symmetry and isotropy arguments.
For the length $l'$ of a rod
$\{p'\in {\Bbb R}^4 \mid x'=0, \;0\le y'\le l,\; z'=0\}$
which is at rest along the $y'$-axis
with respect to the system $K'$ traveling along the $x$-axis
should not depend on the direction of motion; i.e., should only depend
on the absolute magnitude of the velocity.
If this is granted, one obtains
\begin{equation}
\alpha (v) = \sqrt{1-{v^2\over c^2}},
\label{e16-2001-conven}
\end{equation}
and finally the transformation laws
$x\mapsto x'=Lx$   with
\begin{equation}
L(v)=
\left(
\begin{array}{cccc}
{1\over  \sqrt{1-{v^2\over c^2}}}&0&0&-{v\over  \sqrt{1-{v^2\over c^2}}}\\
0&1&0&0\\
0&0&1&0\\
-{v\over c^2 \sqrt{1-{v^2\over c^2}}}&0&0&{1\over  \sqrt{1-{v^2\over
c^2}}}
\end{array}
\right)
\label{e17-2001-conven}
\end{equation}
up to constant translations $a\in {\Bbb R}^4$.
As can be easily checked, $L$ preserves the distance of any two points; i.e.,
$(Lr,Ls)=(r,s)$ for all $r$
and $s$ in ${\Bbb R}^4$.

It would be nice to have a more general result using a more general metric and/or
relaxation of bijectivity.

\section*{Acknowledgments}

I have discussed the ideas exposed here with many researchers in
numerous discussions in Viennese coffee-houses and elsewhere.
I am particularly  indebted to Hans Havlicek.
Almost needless to say, I take the full responsibility
for all errors or misconceptions (if any).
I would like to make it quite clear
that I have not attempted to ``disprove'' relativity theory in
any way. Just on the contrary, these considerations
attempt to extend the domain of validity of relativity theory even
further rather than discredit it.
The extensions deal with universes which are dominated by interactions
which are mediated through velocities different (and maybe also higher
than) the velocity of light.

\end{document}